\documentstyle[psfig]{scienceloc}
\title{Three-Layered Atmospheric Structure in Accretion Disks Around
Stellar-Mass Black Holes}

\author{S.N. Zhang$^{1, 2}$, Wei
Cui$^3$, Wan Chen$^{4, 5}$, Yangsen Yao$^1$, Xiaoling Zhang$^1$, Xuejun
Sun$^1$, Xue-Bing Wu$^6$, Haiguang Xu$^7$ \affiliation{$^1$Physics Department,
University of Alabama in Huntsville, Huntsville, AL 35899}
\affiliation{$^2$SD50, Space Sciences Lab., NASA Marshall Space Flight Center,
Huntsville, AL 35812} \affiliation{$^3$Center for Space Research, MIT,
Cambridge, MA 02139} \affiliation{$^4$Department of Astronomy, University of
Maryland, College Park, MD 20742} \affiliation{$^5$NASA/Goddard Space Flight
Center, Code 661, Greenbelt, MD 20771} \affiliation{$^6$Beijing Astronomical
Observatory, Chinese Academy of Sciences, Beijing 100012, China. P.R.}
\affiliation{$^7$Institute for Space and Astrophysics, Department of Applied
Physics, Shanghai Jiao-Tong University, Shanghai, 200030, China, P.R.}}

\mainauthor{S.N. Zhang}

\summary{Modeling of the x-ray spectra of the Galactic superluminal jet sources
GRS~1915+105 and GRO~J1655-40 reveal a three-layered atmospheric structure in
the inner region of their accretion disks.  Above the cold and optically thick
disk of a temperature 0.2-0.5 keV, there is a warm layer with a temperature of
1.0-1.5 keV and an optical depth around 10. Sometimes there is also a much
hotter, optically thin corona above the warm layer, with a temperature of 100
keV or higher and an optical depth around unity. The structural similarity
between the accretion disks and the solar atmosphere suggest that similar
physical processes may be operating in these different systems.}

\dates{\today}{}

\begin{document}

\maketitle


The sun has a complicated atmosphere, including a photosphere, chromosphere,
transition layer, and an outmost hot
corona\cite{magnetic-field-dynamo,solar-corona}. It is generally thought that
the magnetic activities of the sun may play a significant role in heating the
corona\cite{solar-corona,sturrock-corona-heating}, though other models have
been proposed\cite{scudder}. The atmosphere of the sun is not in hydrodynamical
equilibrium. Consequently, the solar wind is blown outward from the corona.
Coronae and outflows are actually quite common in various types of stellar
environment. Here we present observational evidence for a solar-type atmosphere
for the accretion disks around stellar mass black holes in x-ray
binaries\cite{charles}.

One of the common characteristics of black hole binaries is the so-called
two-component x-ray and gamma-ray spectrum: a soft blackbody-like component at
low energies ($<$ 10 keV) and a hard power-law-like component at high energies
(up to several hundred keV)\cite{black-hole-reviews}. The soft component is
generally attributed to the emission from an optically thick, geometrically
thin cold accretion disk, which is often described by the standard
$\alpha$-disk model\cite{ss-73}. The hard component is attributed to an
optically thin, geometrically thick hot corona in either a plane parallel to
the disk or with a spherical geometry above the disk\cite{liang-pr}. The
prototype models were motivated by the studies of the solar
corona\cite{liang-price-77}.

Recently, more attention has been paid to the similarities between the physical
processes in accretion disks and those in the sun, because the
empirically-invoked viscosity for the disks might have originated from the same
dynamo processes operating on the sun\cite{balbus-hawley}. Consequently, like
in the sun, magnetic turbulence and buoyancy may trigger magnetic flares high
above the disks which could cause intense in situ particle heating and
acceleration, thus powering a disk corona \cite{agn-corona}. The relative
importance of the soft and hard spectral components in black hole binaries is
probably modulated by the energy deposition in the corona by magnetic
flares\cite{fabian-gx339-4} --- more energy deposited in the corona produces a
stronger hard component. It was proposed that there also exists a layer between
the corona and the cold accretion disk, which is directly responsible for the
observed soft component\cite{agn-magnetic-flare,transition-disk}.

To explore the structure of accretion disks in black hole binaries, we carried
out detailed studies of the x-ray spectra of two such sources, GRO~J1655--40
and GRS~1915+105, both of which are galactic superluminal jet
sources\cite{jets}. The higher-than-normal temperature of the soft component of
the two sources is suggested to be caused by the rapid spin of the black holes
in these systems, which results in the disk extending closer to the black hole
horizon\cite{zhang-1655,bh-spin}. Here, we report the results based primarily
on data collected from the Japanese-US x-ray satellite ASCA (Advanced Satellite
for Cosmology and Astrophysics), which has good energy resolution and effective
area in the 0.7-10 keV energy band, ideal for studying the soft component.

For black hole binaries, x-ray emission is powered by the release of
gravitational energy of matter being accreted by the central black holes, which
occurs mostly in the inner region of the accretion disks. Because electron
scattering dominates over free-free absorption in the inner disk\cite{ss-73},
the emergent x-ray spectrum is fully Comptonized (a photon must undergo many
scattering events before escaping). The inner disk region is also radiation
pressure dominated (versus gas pressure), and thus its temperature changes
slowly with radius ($T(r)\propto 1/r^{3/8}$)\cite{liang-pr}, as opposed to a
more rapid temperature variation in the gas pressured dominated disk
($T(r)\propto 1/r^{3/4}$)\cite{liang-pr}. Therefore, we can approximate the
emergent x-ray spectrum from the disk as the Comptonization of a
single-temperature blackbody spectrum by an electron
cloud\cite{comptonization}. In the standard $\alpha$-disk model, the cloud has
the same temperature as the x-ray emission and its optical depth is very large
($>$100)\cite{liang-pr}.

For this analysis, we adopted the Comptonization model by Titarchuk
\cite{comptonization} and fit it with the observed spectra, in order to
determine these key parameters, including the blackbody temperature of the seed
photons (the original photons before the Compton scattering), the electron
temperature and optical depth of the cloud (Table 1). This model fit the data
well for cases where the hard component is negligible. When the hard component
is important, a second Comptonized component is required. Note that although we
use the same Comptonization model for the soft and hard spectral components,
the physical environments for the two are different: the hard component is
produced in an optically thin hot corona (about 100 keV or 10$^9$
 K), while the soft component is produced in an optically thick warm cloud
(about 1 keV or 10$^7$ K). As an example, we plotted the results from spectral
modeling of GRO~J1655-40 and GRS~1915+105 in figure 1.

For the soft component, the results indicate that the temperature of the
electron cloud is higher than the effective temperature of the seed photons to
the Comptonization process , by a factor of 3-6, and the inferred optical depth
of the cloud ($\sim 10$) is much smaller than that expected from the standard
$\alpha$-disk model ($>$100). These results provide observational evidence for
the presence of a lower-density, warm layer outside the standard cold disk. The
hard spectral component, on the other hand, cannot be well constrained by the
ASCA data alone, because the electron temperature of the Comptonizing corona is
higher than the upper end of the ASCA band ($\sim$10 keV). However, the
temperature of the corona can be estimated with the data obtained with the high
energy instruments of CGRO (The US Compton Gamma-Ray Observatory) and RXTE (The
US Rossi-X-ray Timing Explorer). Using the high energy CGRO (20-500 keV) and
RXTE (5-250 keV) data obtained simultaneously with the ASCA observations, we
found that the corona has a temperature of $\sim$100 keV or higher and an
optical depth of the order of unity or less (Table 1). This agrees with
numerous previous reports on the high energy spectra of these two sources, for
example in references\cite{zhang-1655,hard-x-ray,osse}. From Table 1 we can
also see that the temperature of seed photons for the high energy
Comptonization process is very close to that of the warm layer. This implies
that the main source of seed photons feeding the corona is the emergent
Comptonized spectrum from the warm layer outside the cold disk. The
contribution of the hard component to the total x-ray luminosity varies, from
being negligible to being dominant, so the corona is a highly dynamic
environment. At the same time, the warm layer does not seem to vary
appreciably. It is possible that the layer between the cold disk and the hot
corona in Cyg X-1\cite{transition-disk} and the ionized cloud in GRO~J1655-40
and GRS~1915+105 inferred from the iron absorption features\cite{ueda-1655} in
some of the data we used here are in fact the warm layer we identified. However
this warm layer is clearly not produced by the heating of the
corona\cite{transition-disk}, because of the apparent independence between the
relatively stable warm layer and the highly dynamical corona, which sometimes
disappears completely.

Although we used a thermal Comptonization model (the electron kinetic energy is
assumed to follow the Maxwellian distribution) in our fitting, it is worth
pointing out that non-thermal electron energy distribution (for the high energy
spectral component) may also be consistent to the data, as implied by steep
power-law spectra extending beyond several hundred keV in some observations
\cite{osse}. In this case, the electron temperature inferred in our model
fitting should be considered the lower limit to the kinetic energy of electrons
in the corona. Therefore the corona may also be in the form of jets/outflow
from the black hole or near-spherical converging flow into the black
hole\cite{converging-flow}. Current x-ray data alone do not allow us to
distinguish these different corona geometry unambiguously.

The inferred structure of accretion disks of black hole binaries can be
compared to the structure of the solar atmosphere (Fig. 2). The photosphere,
chromosphere and corona of the sun appear to correspond to the surface of the
cold disk, the overlaying optically thick warm layer, and the optically thin
hot corona, respectively. The transition region between the chromosphere and
the corona in the solar atmosphere, however, could not be identified in disks
from our current spectral fitting. It cannot correspond to the warm layer,
because the latter is observed even in the absence of the corona (the hard
component). The temperatures of the three regions are higher in the accretion
disks by approximately a factor of 500 than the corresponding regions in the
sun. This supports the notion that magnetic activity is responsible for
powering the upper atmosphere in both cases, giving $T\propto E^{1/4}\propto
B^{1/2}$ and thus $T_{\rm DISK}/T_{\rm SUN}\approx (B_{\rm DISK}/B_{\rm
SUN})^{1/2}\approx (10^8G/500G)^{1/2}\approx 500$.

Because the solar wind is driven out primarily by strong coronal activities, by
analogy, we argue that the corona surrounding accretion disks of black hole
binaries is also a source of outflow. In fact a recent accretion disk model
suggests that magnetic field driven jets and outflow are also important in the
angular momentum transfer, which is essential in order for the accretion
process to operate in these systems\cite{outflow-adaf}. The connection between
corona and outflow is also supported by the fact that radio emissions from
black hole binaries seems always accompanied with the detection of significant
hard x-ray emission\cite{hard-x-ray}. Although the magnetic fields may be
generated by the dynamo processes as a result of the differential rotation in
both the sun and the accretion disks of black hole
binaries\cite{magnetic-field-dynamo}, the two types of systems are different.
In the sun, the source of radiation energy is the nuclear burning in the core,
and only a small portion of the total energy is converted to the magnetic
field. In accretion disks, however, all radiation energy comes from the viscous
dissipation of gravitational energy, which may originate in magnetic
turbulence\cite{balbus-hawley}. It is, therefore, natural that the magnetic
field related energy dissipation in accretion disks is more important than that
in the sun. Consequently our results provide strong support to theoretical
predictions that relativistic jets and outflow from these systems are
magnetic-field driven\cite{blandford-jets}. In fact it has been realized
recently that there might exist a `magnetic switch' in these systems; when the
magnetic field activity exceeds a certain limit, fast and relativistic jets may
be produced\cite{meier}. Our results reported here and the fact that both
GRO~J1655-40 and GRS~1915+105 have been observed to produce highly relativistic
jets\cite{jets} provide support to this magnetic switch theory. It is worth
noting that disk coronae powered by magnetic flares are also believed to exist
in the accretion disks around supermassive black holes\cite{agn-corona}.
Therefore, similar physical processes may operate in systems with different
properties and scales.
\begin{figure*}
\hbox{

\psfig{figure=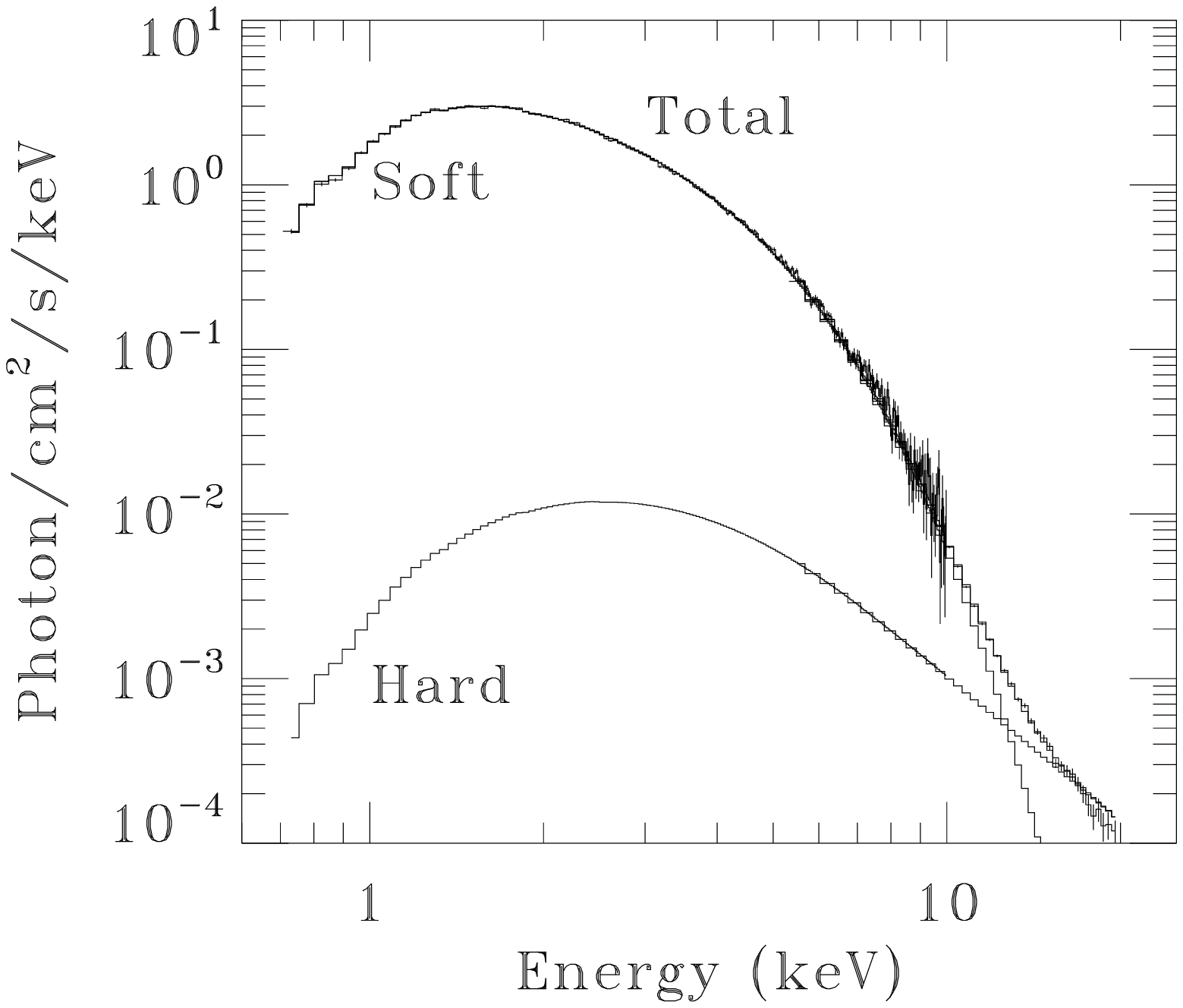,width=3.4in}
\psfig{figure=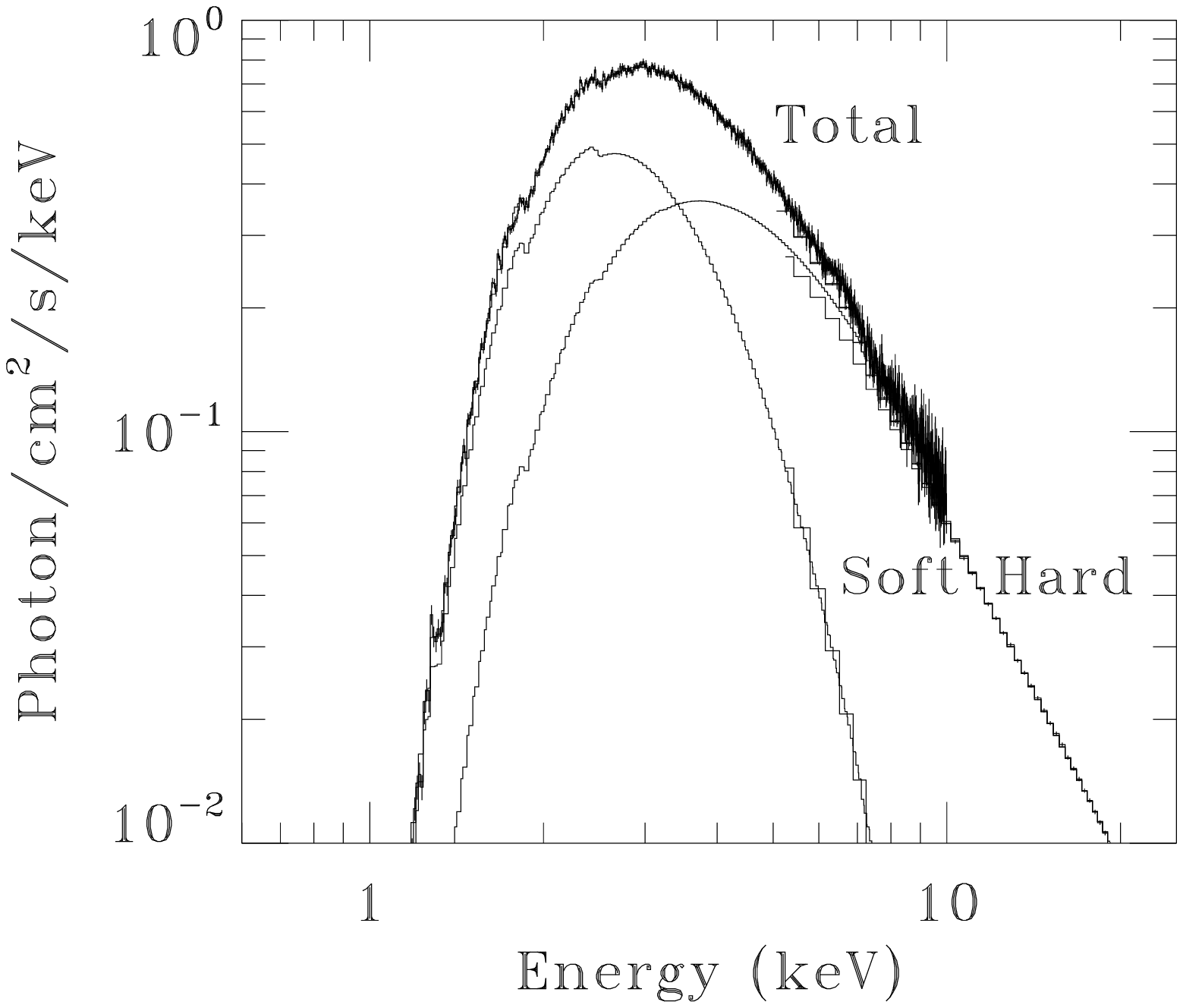,width=3.4in} }

\caption{Sample spectra of GRO~J1655-40 ({\it left}: observed on 97/02/26) and
GRS~1915+105 ({\it right:} observed on 96/10/23). Both soft and hard components
are important in these spectra. ASCA data: 0.7-10 keV; RXTE data: 5-20 keV
(RXTE data above 20 keV not shown for clarity of the figures).}

\end{figure*}

\begin{figure*}
\hbox{ \psfig{figure=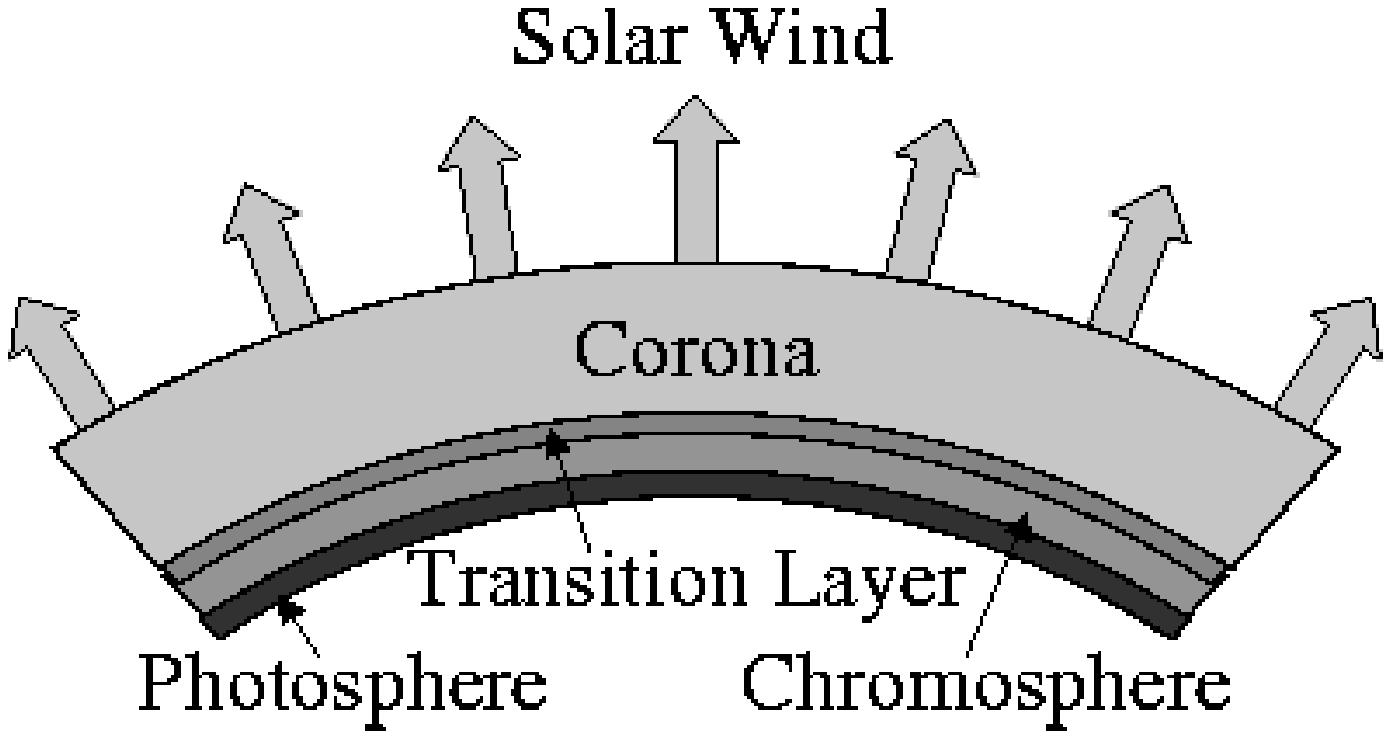,width=3.4in,angle=270}
\psfig{figure=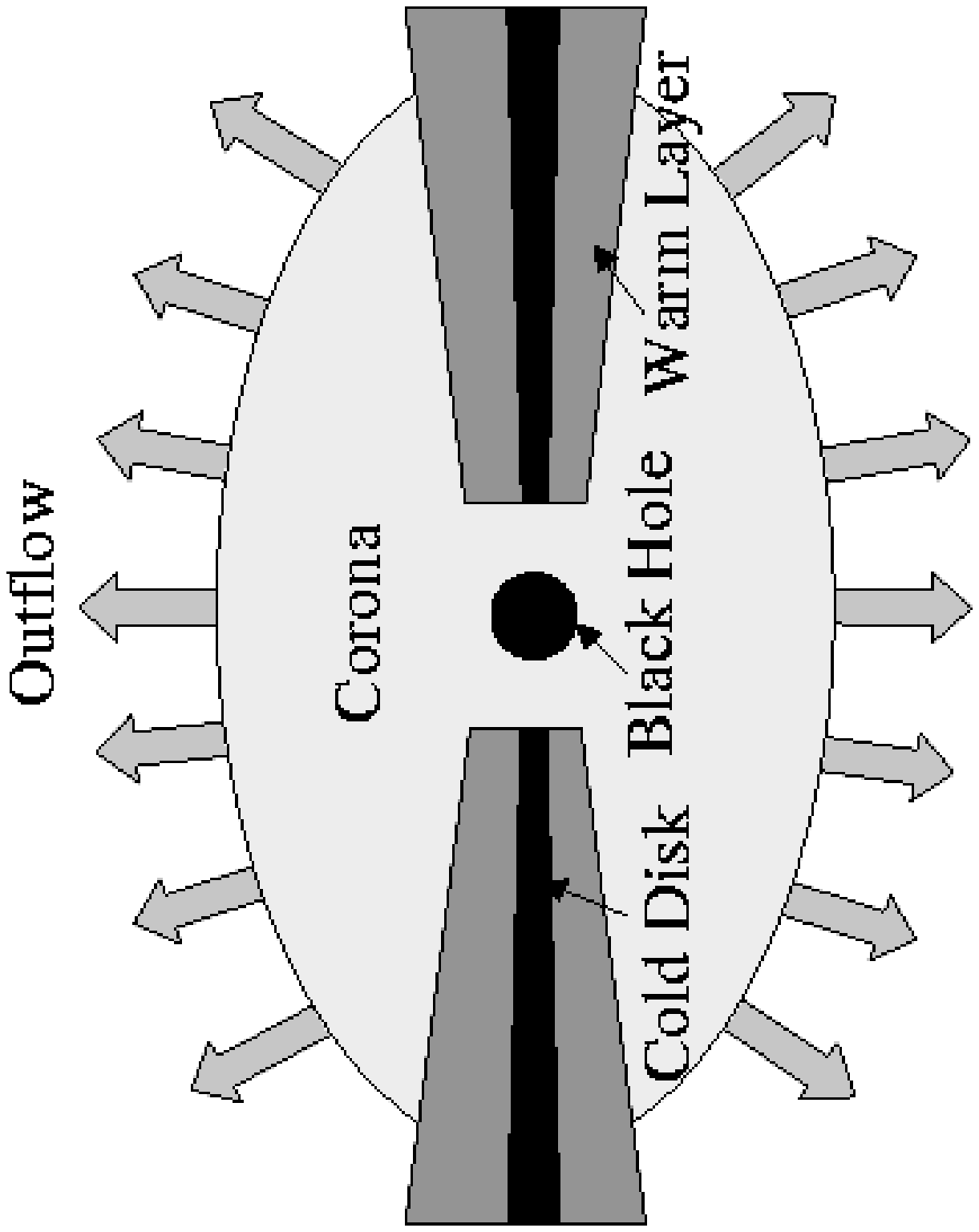,width=3.3in,angle=270} }
 \caption{Schematic diagrams of the solar atmosphere and accretion
disk structure. The temperatures in the solar atmosphere are approximately:
6$\times$10$^3$ K (photosphere),  3$\times$10$^4$ K (chromosphere), and
2$\times$10$^6$ K (corona), respectively. For the black hole disk atmosphere,
the corresponding temperatures are approximately 500 times higher:
3$\times$10$^6$ K (cold disk), 1.5$\times$10$^7$ K (warm layer), and
1$\times$10$^9$ K (corona), respectively.}
\end{figure*}

\begin{table}
\begin{tabular}{@{}c@{}c@{}c@{ }c@{ }c@{ }c@{ }c@{ }c@{ }c@{ }c@{}}
\hline

Observation & N$_{\rm H}$ & \multicolumn{4}{c}{Soft Component} &
\multicolumn{4}{c}{Hard Component}\\

& 10$^{22}$ & {kT$_0$} & {kT$_{\rm e}$} & {$\tau$} & {F$_{\rm x}$} &
 {kT$_0$} & {kT$_{\rm e}$} &
{$\tau$} & {F$_{\rm x}$}\\

(1) & H/cm$^{2}$ & keV & keV & & (2) & keV & keV & & (2) \\ \hline

A/94/08/23 & 0.26 & 0.20 & 1.56 & 8.1 & 0.32 \\

A/94/09/27 & 0.39 & 0.21 & 1.21 & 12.3 &2.1 \\

A/95/08/15 & 0.57 & 0.33 & 1.02 & 12.8 & 5.9  & 1.31 & $>$95 & $<$0.01 & 1.7 \\

A/97/02/26 & 0.56 & 0.26 & 0.94 & 14.4 & 4.8  & 0.92 & $>$47 & $<$1.0 & 0.1\\

B/94/09/27 & 3.45 & 0.33 & 1.31 & 12.7 & 1.8  & \\

B/95/04/20 & 3.55 & 0.51 & 1.42 & 13.8 & 3.4  \\

B/96/10/23 & 3.82 & 0.46 & 1.10 & 8.1 & 2.0 & 1.06 & $>$87 & $<0.2$ & 2.9 \\

B/97/04/25 & 3.60 & 0.46 & 1.11 & 8.9 &1.0 & 1.07 & $>$35 & $<$0.8 & 1.4\\

B/98/04/04 & 3.57 & 0.45 & 1.06 & 9.7 & 1.0 & 1.12 &$>$26 & $<1.0$ & 1.3\\
\hline
\end{tabular}
\caption{Results for all ASCA observations of GRO~J1655-40 and GRS~1915+105;
simultaneous CGRO and RXTE data are also used for determining the high energy
component above 10 keV. In about half of the observations the high energy
component is detectable or dominating the total luminosity. Notes: (1)
A~=~GRO~J1655-40 and B~=~GRS~1915+105; (2) the un-absorbed flux is in units of
10$^{-8}$ erg/cm$^{2}$/s. N$_{\rm H}$: interstellar neutral hydrogen column
density; kT$_{0}$: seed-photon temperature to the Comptonization process;
kT$_{\rm e}$: electron temperature of the Comptonization medium; $\tau$:
optical depth of the Comptonization medium.}
\end{table}

\bibliographystyle{science}

\smallskip
\noindent {\small {\bf Acknowledgements.} We thank Drs. Junhan You of SJTU
(China), Robert Shelton of UAH, Alan Harmon of NASA/MSFC, Kajal Ghosh of
NRC/MSFC and Lev Titarchuk of GSFC for useful discussions, and Ken Ebisawa of
USRA/GSFC for help on ASCA data analysis. We acknowledge partial financial
support from NASA GSFC under the Long Term Space Astrophysics program and
several guest investigations, and from NASA MSFC through contract NCC8-65.
\medskip
\noindent {\small Correspondence should be addressed to S.N.Z. (e-mail:
zhangsn@email.uah.edu).}
\end{document}